\documentstyle[12pt,a4,epsfig]{article}
\begin{document} 
\vspace{-1.5cm}
\rightline{\vbox{ \hbox{PNU-NTG-05/98}}}
\vspace{0.5cm}

\begin{center} 
{\large \bf Effective $\Delta S=1$ weak chiral Lagrangian 
\\ from the instanton-induced chiral quark model}
\footnote{Talk given at the 1998 YITP Workshop on QCD and 
Hadron Physics, 14-16 October, Kyoto, Japan.}  \\[0.5cm] 
Mario Franz$^{(1)}$, Hyun-Chul Kim$^{(2)}$ , and 
Klaus Goeke$^{(1)}$  \\ [0.3cm] 
$^1$Institute for Theoretical  Physics  II,   P.O. Box 102148,  \\  
Ruhr-University Bochum, 
 D-44780 Bochum,   Germany \\ [0.3cm]
$^2$ Department of Physics, Pusan National University,\\ 
609-735 Pusan, Republic of Korea [0.3cm] 
\end{center}
 
\begin{abstract} 
We present the recent investigation of 
the $\Delta S = 1$ effective weak chiral Lagrangian 
within the framework of the instanton-induced chiral quark model.  
Starting from the effective four-quark operators, 
we derive the effective weak chiral action by integrating out the 
constituent quark fields.  Employing the derivative expansion, 
we obtain the effective weak chiral Lagrangian to ${\cal O} (p^4)$ 
order with low energy constants.
\end{abstract} 

\date{December, 1998} 

Chiral perturbation theory ($\chi$PT), known as an effective field
theory in very low-energy regime, explains low-energy phenomena
in strong interactions very well.  Based on its success in describing strong 
interactions,  $\chi$PT was also applied to the nonleptonic processes
\cite{Kamboretal,Esposito}.  However, the effective weak chiral Lagrangian
of order ${\cal O}(p^4)$ brings in too many low energy constants (LECs) 
to be determined by experimental data, as the chiral corrections are 
considered.  In order to proceed to make numerical calculations,
it is inevitable to rely on models which provide the LECs.  

Recently, Antonelli {\em et al.}~\cite{Antonellietal} 
applied the $\chi$QM in order to obtain the effective weak chiral 
Lagrangian in the lowest order (${\cal O} (p^2)$) in the chiral expansion.
The large $N_c$ expansion was also considered in orders 
${\cal O} (N^{2}_c)$, ${\cal O} (N_c)$, and 
${\cal O} (\alpha_s N_c)$.  In their study, the correction 
from order ${\cal O} (\alpha_s N_c)$ plays an essential role in reaching
the agreement with phenomenology.  However, one should mention that
treating order ${\cal O} (\alpha_s N_c)$ requires a special care. 
As noted by Ref.~\cite{PichRafael}, ${\cal O} ([\alpha_s N_c]^2)$ is
in fact the same order as ${\cal O} (\alpha_s N_c)$ and might be 
nonnegligible.  

More recently, Ref.~\cite{Bertolinietal} extended 
the work of Ref.~\cite{Antonellietal} to the next-to-leading order
in the chiral expansion.  However, they calculated directly the transition
amplitudes of the decay $K\rightarrow \pi\pi$ to order ${\cal O}(p^4)$,
which is in a sense different from the original scheme of chiral
perturbation theory.

In this work we rather want to furnish the LECs not only to 
order ${\cal O}(p^2)$ but also to order ${\cal O} (p^4)$ within the 
framework of the instanton-induced chiral quark model.   

The effective chiral action  with the $\Delta S = 1$ effective weak
Hamiltonian can be written as follows:
\begin{equation}
\exp{\left(- S^{\Delta S = 1}_{\rm eff}\right)} \;=\; 
\int {\cal D} \psi {\cal D} \psi^\dagger 
\exp \left[\int d^4 x \left(\psi^{\dagger} D
\psi - {\cal H}^{\Delta S = 1}_{\rm eff}\right)
\right],
\label{Eq:partw}
\end{equation} 
where $D$ is the Dirac operator with the pseudo-Goldstone 
boson~\cite{DP}
\begin{equation}
D \;=\;
i\rlap{/}{\partial} + iM U^{\gamma_5}.
\end{equation}
The constituent quark mass $M$ is a momentum-dependent mass which 
characterizes the instanton-induced chiral quark model.  The effect 
of the momentum dependence of the $M$  will be considered
in a later work~\cite{FKG}.  
$U^{\gamma_5}$ denotes the chiral meson field.  The effective weak 
quark Hamiltonian 
${\cal H}^{\Delta S = 1}_{\rm eff}$ consists of ten four-quark operators 
among which only seven operators are independent:
\begin{equation}
{\cal H}^{\Delta S = 1}_{\rm eff}
 \;=\; -\frac{G_F}{\sqrt{2}} V_{ud} V^*_{us}
\sum_i c_i (\mu) {\cal Q}_i (\mu) + {\rm h.c.} .
\end{equation} 
The Hamiltonian ${\cal H}^{\Delta S = 1}_{\rm eff}$ 
can be found elsewhere (for example, see Ref.~\cite{Burasetal}).  Since 
the Fermi constant $G_F$ is very
small, we can expand Eq.(\ref{Eq:partw}) with regard to it and
obtain the following expression:
\begin{equation}
{\cal L}^{\Delta S = 1}_{\rm eff} \;=\; -\frac{1}{\cal N}
\int {\cal D} \psi {\cal D} 
\psi^\dagger {\cal H}^{\Delta S = 1}_{\rm eff} \exp  
\left[\int d^4 x \psi^\dagger D \psi\right].
\end{equation}
By integrating out the quark fields and making the derivative expansion,
we derive the effective chiral Lagrangian as follows:
\begin{eqnarray}
{\cal L}^{\Delta S = 1,{\cal O} (p^2)}_{\rm eff}&=& 
-\frac{G_F}{\sqrt{2}} V_{ud} V^*_{us} f^4_{\pi} 
\left[g_{\underline{8}} \left\langle\lambda^2_{3} L_\mu L^\mu \right\rangle
\right. \nonumber \\ && + \left.g_{\underline{27}} \left(\frac23
\left\langle \lambda^1_{2} L_\mu\right\rangle
\left\langle \lambda^3_{1} L^\mu \right\rangle  + 
\left\langle \lambda^3_{2} L_\mu\right\rangle
\left\langle \lambda^1_{1} L^\mu \right\rangle \right) \right]
\;+\; {\rm h.c.}  \nonumber \\ 
&=& {\cal L}^{(1/2)}_{\underline{8}} \;+\; \frac19 
{\cal L}^{(1/2)}_{\underline{27}} \;+\;  
\frac59 {\cal L}^{(3/2)}_{\underline{27}},
\label{Eq:Lcpt2}
\end{eqnarray}
where
\begin{eqnarray}
\label{Eq:L8L27}
{\cal L}^{(1/2)}_{\underline{8}} &=& 
-\frac{G_F}{\sqrt{2}} V_{ud} V^*_{us}f^4_{\pi} 
g_{\underline{8}} \left\langle\lambda^2_{3} 
L_\mu L^\mu \right\rangle\;+\; {\rm h.c.}, \\
{\cal L}^{(1/2)}_{\underline{27}} &=& 
-\frac{G_F}{\sqrt{2}} V_{ud} V^*_{us}f^4_{\pi}
g_{\underline{27}}\left(\left\langle \lambda^1_{2} L_\mu\right\rangle
\left\langle \lambda^3_{1} L^\mu \right\rangle
 -
\left\langle \lambda^3_{2} L_\mu\right\rangle
\left\langle \lambda^1_{1} L^\mu \right\rangle 
- 5 \left\langle \lambda^3_{2} L_\mu\right\rangle
\left\langle \lambda^3_{3} L^\mu \right\rangle\right) +\; {\rm h.c.}, 
\nonumber \\
{\cal L}^{(3/2)}_{\underline{27}} &=& 
-\frac{G_F}{\sqrt{2}} V_{ud} V^*_{us}f^4_{\pi}
g_{\underline{27}}\left(\left\langle \lambda^1_{2} L_\mu\right\rangle
\left\langle \lambda^3_{1} L^\mu \right\rangle 
+2 \left\langle \lambda^3_{2} L_\mu\right\rangle
\left\langle \lambda^1_{1} L^\mu \right\rangle 
+ \left\langle \lambda^3_{2} L_\mu\right\rangle
\left\langle \lambda^3_{3} L^\mu \right\rangle\right) +\; {\rm h.c.}
\nonumber
\end{eqnarray}
Note that the isospin $T=1/2$ part of the eikosiheptaplet 
is suppressed by the prefactor, so that one can treat the eikosiheptaplet 
part of the Lagrangian as a pure $T=3/2$ one.      
$g_{\underline{8}}$ and $g_{\underline{27}}$ can be extracted from the 
$K\rightarrow \pi\pi$ decay rate.  At the tree level, the $\Delta T = 1/2$ 
enhancement is reflected in these constants.

We first revisit the recent results of Ref.~\cite{Antonellietal} but
we concentrate only on the low energy constants.  To obtain the comparable
results to the empirical data, Ref.~\cite{Antonellietal}   
considered the effect of the gluon condensates coming from the external
backgroud gluon fields~\cite{PichRafael} in order 
${\cal O}(\alpha_s N_c)$.  With orders 
${\cal O}(N^2_{c})$, ${\cal O} (N_c)$ and 
${\cal O}(\alpha_s N_c)$ taken into account, one arrives at
following results in the leading order ${\cal O} (p^2)$:
\begin{eqnarray}
 g_{\underline{8}}^{\delta_{\langle GG \rangle}}
&=&
 \left(- \frac{2}{5} + {1 \over N_c} ( 1 - \delta_{\langle GG \rangle}) 
\frac{3}{5} \right) c_1
+ \left(  \frac{3}{5} - {1 \over N_c} ( 1 - \delta_{\langle GG \rangle}) 
\frac{2}{5} \right) c_2
+ {1 \over N_c} ( 1 - \delta_{\langle GG \rangle}) c_3 + c_4
\nonumber \\
&& \hspace{5mm} + \left( {\langle \bar{q} q \rangle \over N_c f_\pi^2 M} -
{ \langle \bar{q} q \rangle M \over 8 f_\pi^4 \pi^2}
\right) c_5
+ \left( { \langle \bar{q} q \rangle \over f_\pi^2 M} -
{N_c \langle \bar{q} q \rangle M \over 8 f_\pi^4 \pi^2}
 \right) c_6
\nonumber \\
&& \hspace{5mm} + \left( - \frac{3}{5} + {1 \over N_c} 
( 1 - \delta_{\langle GG \rangle}) \frac{2}{5} \right) c_9
+ \left(  \frac{2}{5} - {1 \over N_c} ( 1 - \delta_{\langle GG \rangle}) 
\frac{3}{5} \right) c_{10}
\\
g_{\underline{27}}^{\delta_{\langle GG \rangle}}
&=&
\left( 1 + {1 \over N_c} ( 1 - \delta_{\langle GG \rangle}) \right) 
\left(
\frac{3}{5} c_1 + \frac{3}{5} c_2 + \frac{9}{10} c_9 + \frac{9}{10} c_{10}
\right),
\end{eqnarray}
where
\begin{equation}
\delta_{\langle GG\rangle} = \frac{N_c}{2} 
\frac{\langle \alpha_s GG /\pi \rangle}
{16\pi^2 f^4_{\pi}}.
\end{equation}

Figure draws the dependence of the $g_{\underline{8}}/g_{\underline{27}}$
on the gluon condensate.  

\vspace{0.5cm}
\centerline{\epsfig{file=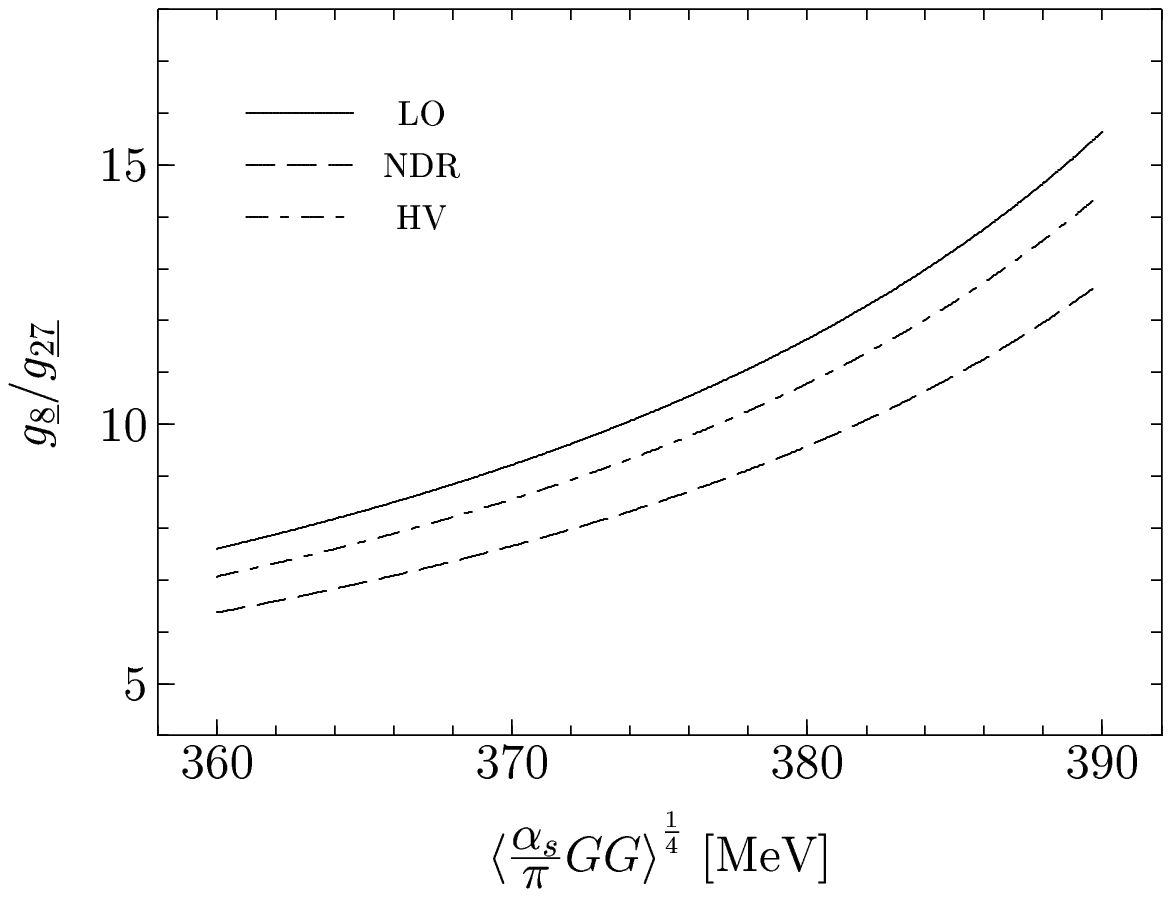,height=7.0cm,width=10.0cm,silent=}}
\begin{center}
\parbox{14cm}{\footnotesize {\bf Fig.3}: Dependence of the
$g_{\underline{8}}/g_{\underline{27}}$ on the gluon condensate 
$\langle {\alpha_s \over \pi} GG \rangle$.
The solid curve denotes the LO renormalization scheme in 
Ref.\protect{\cite{Burasetal}}, while the dashed curve and dot-dashed one
stand for the NDR and the HV schemes, respectively.  The value of
the constituent quark mass $M =300 \; {\rm MeV}$ is used and the quark 
condensate $\langle \bar{q}q\rangle/2 =-(250\; {\rm MeV})^3$ 
is employed.}
\vspace{0.8cm}
\end{center}

\noindent
As the gluon condensate increases, the 
ratio $g_{\underline{8}}/g_{\underline{27}}$ is improved remarkably, 
so that we are able to obtain the reasonable value of the ratio 
around $(380\;\mbox{MeV})^4$, which is acceptable with the counterterms
being considered~\cite{KMW}.   However, there is one important caveat.  As 
was mentioned in Ref.~\cite{PichRafael}, the ${\cal O} (\alpha N_c)^2$
corrections which are of order ${\cal O} (1)$ as well were neglected,
based on the argument that they involve condensates of higher dimension
requiring higher powers of the normalization scale.    
However, keeping in mind the fact that the ${\cal O} (\alpha_s N_c)$ 
corrections enhance the result of the ratio $g_{\underline{8}}
/g_{\underline{27}}$ so radically as shown above, the higher-order
terms might give some contribution.  We would rather regard the above results
as a mere indication of the importance of gluon effects in the 
nonleptonic processes.
 
An effective weak chiral Lagrangian to order ${\cal O}(p^4)$ in $\chi$PT 
with a minimal set  of independent terms was given by  
Ecker {\em et al.}~\cite{EKW} and Esposito-Far\`ese~\cite{Esposito} 
(presented in Minkowski space).  A part of the corresponding low energy
constants are obtained from the present study:
\begin{eqnarray}
{\cal L}^{\Delta S=1,{\cal O}(p^4)}_{\rm eff} 
&=& - {G_F \over \sqrt{2}} V_{\rm ud} V_{\rm us}^* 
f_\pi^2 \left[ \left(N_{1}^{(\underline{8})} 
        \langle \lambda_6  L_\mu  L^\mu  L_\nu L^\nu  \rangle
\; + \; N_{2}^{(\underline{8})} \cdot
        \langle \lambda_6  L_\mu  L^\nu  L_\nu L^\mu  \rangle
\right.\right. \nonumber \\
 && \hspace{2.5cm}+\; N_{3}^{(\underline{8})}
 \langle \lambda_6  L_\mu  L_\nu \rangle \langle  L^\mu L^\nu  \rangle
\;+ \; N_{4}^{(\underline{8})} 
\langle \lambda_6  L_\mu \rangle \langle  L^\mu  L_\nu L^\nu  \rangle
\nonumber \\
&&\hspace{2.5cm}  +\; \left.N_{28}^{(\underline{8})}   i \,
        \epsilon_{\mu \nu \rho \delta}
        \langle \lambda_6  L^\mu \rangle
        \langle  L^\nu  L^\rho  L^\delta  \rangle \right)\nonumber \\
&+& \frac59 {t}_{ijkl} \, \left( N_{1}^{(\underline{27})} 
        \langle \lambda_{ij} L_\mu L^\mu \rangle
        \langle \lambda_{kl} L_\nu L^\nu \rangle\right.
\;+\; N_{2}^{(\underline{27})}   
       \langle \lambda_{ij} L_\mu L_\nu \rangle
        \langle \lambda_{kl} L^\mu L^\nu \rangle
\nonumber \\
&&\hspace{0.2cm} +\;  N_{3}^{(\underline{27})}  
        \langle \lambda_{ij} L_\mu L_\nu \rangle
        \langle \lambda_{kl}  L^\nu L^\mu \rangle
\; +\; N_{4}^{(\underline{27})}   
        \langle \lambda_{ij} L_\mu \rangle
        \langle \lambda_{kl} L_\nu L^\mu  L^\nu \rangle
\nonumber \\
&&\hspace{0.2cm} +\; N_{5}^{(\underline{27})}   
    \langle \lambda_{ij} L_\mu \rangle
    \langle \lambda_{kl} \left\{ L^\mu ,  L_\nu  L^\nu \right\} \rangle
\;+\; N_{6}^{(\underline{27})}  
        \langle L_\mu L^\mu \rangle
        \langle \lambda_{ij} L_\nu \rangle
        \langle \lambda_{kl} L^\nu \rangle
\nonumber \\ 
&& \hspace{0.2cm} \left.\left. +\; N_{20}^{(\underline{27})}   
 i \,        
        \epsilon_{\mu \nu \rho \delta}
        \langle \lambda_{ij} L^\mu L^\nu  \rangle
        \langle \lambda_{kl} L^\rho L^\delta \rangle
\;+\; N_{21}^{(\underline{27})}    i \,
        \epsilon_{\mu \nu \rho \delta}
        \langle \lambda_{ij} L^\mu \rangle
        \langle \lambda_{kl} L^\nu L^\rho L^\delta  \rangle
\right) \right].
\end{eqnarray}
As noted by G. Ecker {\em et al.}~\cite{EKW}, the LECs 
$N_{1}^{(\underline{8})}$, $N_{2}^{(\underline{8})}$,
$N_{3}^{(\underline{8})}$, and $N_{4}^{(\underline{8})}$ 
contribute to the process $K\rightarrow 3\pi$ while
$N_{28}^{(\underline{8})}$ does to the radiative $K$-decays.  In particular,
the $N_{28}^{(\underline{8})}$ is related to the chiral 
anomaly~\cite{BEP}.  Note that all LECs in the eikosiheptaplet are 
degenerate.  

Taking into account the $N_c$ corrections, 
the LECs are derived as follows:
\begin{eqnarray}
N_{1}^{(\underline{8})} &=&
\left( -{N_c^2 M^2 \over 128 \pi^4 f_\pi^2 } + {N_c \over 8 \pi^2}
- {f_\pi^2 \over 2 M^2} \right) c_6  \;+\;
\left( -{N_c M^2 \over 128 \pi^4 f_\pi^2 } + {1 \over 8 \pi^2}
- {f_\pi^2 \over 2 N_c M^2} 
\right) c_5,
\\
N_{2}^{(\underline{8})} &=&
{N_c \over 60 \pi^2}
\left( \left( -2 +{1 \over N_c} 3 \right) c_1 + 
\left( 3 - {1 \over N_c} 2 \right) c_2 
+ {1 \over N_c} 5 c_3 + 5 c_4 \right. \nonumber \\
&& + \; \left.
\left( - 3 + {1 \over N_c} 2 \right) c_9 + 
\left( 2 - {1 \over N_c} 3 \right) c_{10} \right),
\\
N_{3}^{(\underline{8})} &=&
0,
\\
N_{4}^{(\underline{8})} &=&
{N_c \over 60 \pi^2 }
\left( \left( \frac{3}{2} - {1 \over N_c} \right) c_1 
+ \left( - 1 + {1 \over N_c} \frac{3}{2} \right) c_2 
+ \frac{5}{2} c_3 + {1 \over N_c} \frac{5}{2} c_4 
\right. \nonumber \\ && \hspace{10mm} \left.
- \frac{5}{2} c_5 - {1 \over N_c} \frac{5}{2} c_6
+ \left( 1 - {1 \over N_c} \frac{3}{2} \right) c_9 
+ \left( -  \frac{3}{2} + {1 \over N_c} \right) c_{10} \right),
\\
N_{28}^{(\underline{8})} &=&
{N_c \over 60 \pi^2}
\left(  \left( -\frac{3}{2} + {1 \over N_c} \right) c_1 
+ \left( 1 - {1 \over N_c}  \frac{3}{2} \right)  c_2 
- \frac{5}{2} c_3 -  {1 \over N_c} \frac{5}{2} c_4 
\right. \nonumber \\ && \hspace{10mm} \left.
- \frac{5}{2} c_5 - {1 \over N_c} \frac{5}{2} c_6 
+ \left( - 1 + {1 \over N_c} \frac{3}{2} \right) c_9 
+ \left(  \frac{3}{2} - {1 \over N_c} \right) c_{10} \right),
\\ 
N_{1}^{(\underline{27})} &=&
N_{5}^{(\underline{27})} =
N_{6}^{(\underline{27})} =
N_{20}^{(\underline{27})} = 0,
\\
N_{2}^{(\underline{27})} &=&
-N_{3}^{(\underline{27})} =
-N_{4}^{(\underline{27})} =
N_{21}^{(\underline{27})} =
{N_c \over 60 \pi^2 } \left( 1 + {1 \over N_c} \right) 
\left( - 3 c_1 - 3 c_2 - \frac{9}{2} c_9 - \frac{9}{2} c_{10} \right).
\label{Eq:LECs}
\end{eqnarray}
The corresponding numerical results and further discussions such as the 
contribution of the quark axial-vector constant $g_A$ and the 
momentum-dependence of the constituent quark mass will appear 
soon~\cite{FKG}.

In summary, we have determined a part of the LECs for 
the ${\cal O}(p^4)$ and discussed the relevance of the gluon condensate
to the LECs in the leading order.  
  
The work is supported in part by COSY, DFG, and BMBF.  HCK acknowledges 
the financial support of the Korean Research Foundation
made in the program year of 1998.

\end{document}